\documentclass[twocolumn,nodate,showpacs,preprintnumbers,amsmath,amssymb,aps,prd]{revtex4}
\usepackage{graphicx}
\usepackage{color}
\usepackage{amsmath,amssymb,amsfonts,amsthm,latexsym}

\newcommand*{\Z}{{\mathbb Z}}

\allowdisplaybreaks

\def\f{\frac}

\begin{document}
\preprint{AEI-2009-042}
\title{Alternative quantization of the Hamiltonian in loop quantum cosmology II: \\
Including the Lorentz term}
\author{Jinsong Yang${}^1$}
\author{You Ding${}^{1,2}$}
\author{Yongge Ma${}^{1,3}$}
\email{mayg@bnu.edu.cn} \affiliation{\small{${}^1$Department of
Physics, Beijing Normal
University, Beijing 100875, China}\\
\small{${}^{2}$ Centre de Physique Th\'eorique de Luminy,
Universit\'e de la M\'editerran\'ee, F-13288 Marseille, EU}\\
\small{${}^{3}$ Max Plank Institute for Gravitational Physics (AEI),
Am M\"{u}hlenberg 1, 14476 Potsdam, Germany}}

\begin{abstract}
Since there are quantization ambiguities in constructing the
Hamiltonian constraint operator in isotropic loop quantum cosmology,
it is crucial to check whether the key features of loop quantum
cosmology are robust against the ambiguities. In this paper, we
quantize the Lorentz term of the gravitational Hamiltonian
constraint in the spatially flat FRW model by two approaches
different from that of the Euclidean term. One of the approaches is
very similar to the treatment of the Lorentz part of Hamiltonian in
loop quantum gravity and hence inherits more features from the full
theory. Two symmetric Hamiltonian constraint operators are
constructed respectively in the improved scheme. Both of them are
shown to have the correct classical limit by the semiclassical
analysis. In the loop quantum cosmological model with a massless
scalar field, the effective Hamiltonians and Friedmann equations are
derived. It turns out that the classical big bang is again replaced
by a quantum bounce in both cases. Moreover, there are still great
possibilities for the expanding universe to recollapse due to the
quantum gravity effect.
\end{abstract}
\maketitle

\section{Introduction}
An important motivation of the theoretical search for a quantum
theory of gravity is the expectation that the singularities
predicted by classical general relativity would be resolved by the
quantum gravity theory. This expectation has been confirmed by the
recent study of certain isotropic models in loop quantum cosmology
(LQC) \cite{bojowald,aps1,aps}, which is a simplified
symmetry-reduced model of a full background-independent quantum
theory of gravity \cite{lrr}, known as loop quantum gravity (LQG)
\cite{al,rov,thiem,hhm}. In loop quantum cosmological scenario for a
universe filled with a massless scalar field, the classical
singularity gets replaced by a quantum bounce
\cite{aps,bojow-r,ash-r}. Moreover, It is also revealed in the effective
scenarios that there are great possibilities for a spatially flat
FRW expanding universe to recollapse due to the quantum gravity
effect \cite{dmy1}. However, as in the ordinary quantization
procedure, there are quantization ambiguities in constructing the
Hamiltonian constraint operator. Thus it is crucial to check the robustness of the
key results against the quantization ambiguities. Moreover, since LQC serves as a simple
arena to test ideas and constructions induced in
the full LQG, it is important to implement those treatments from the full theory to LQC as more
as possible.

In the previous paper \cite{dmy2} of this series of work, it has
been shown that the above key features of LQC are robust against a
quantum ambiguity arising from the quantization of the field
strength of the gravitational connection. In this paper we will
propose alternative versions of Hamiltonian operator for isotropic
LQC, which inherit more features from full LQG comparing to the
conventional one so far considered in the literatures. It is well
known that the Hamiltonian constraint in the full theory is composed
of two terms, the so-called Euclidean and Lorentz terms. In
spatially flat and homogeneous models, the two terms are
proportional to each other. Thus people usually rewrite the Lorentz
term in the form of the Euclidean one classically and then quantize
their combination \cite{aps}. However, this treatment is impossible
in the full theory, where the Lorentz term has to be quantized in a
form quite different from the Euclidean one \cite{thiem,ThiemQ}. The
issue that we are considering is what would happen in the improved
dynamics setting of LQC if one kept the distinction of the two terms
as in full theory rather than mixed them. Could one construct an
operator corresponding to the Lorentz term in a way similar to that
in full LQG? If so, could the classical big bang singularity still
be replaced by a quantum bounce in the new quantum dynamics? To
answer these questions, two alternative Hamiltonian constraint
operators including the Lorentz terms are constructed respectively
in the improved scheme in this paper, which are both shown to have
the correct classical limit by the semiclassical analysis. In the
spatially flat FRW model with a massless scalar field, the effective
Hamiltonians and Friedmann equations are derived in both case. It
turns out that the classical big bang is again replaced by a quantum
bounce. Moreover, there are still great possibilities for the
expanding universe to recollapse due to the quantum gravity effect.

In the spatially flat and isotropic model, one has to first
introduce an elementary cell ${\cal V}$ and restrict all
integrations to this cell. Fix a fiducial flat metric
${{}^o\!q}_{ab}$ and denote by $V_o$ the volume of the elementary
cell ${\cal V}$ in this geometry. The gravitational phase space
variables
---the connections $A_a^i$ and the density-weighted triads
$E^a_i$--- can be expressed as \cite{math}
\begin{align}
A_a^i = c\, V_o^{-1/3}\,\, {}^o\!\omega_a^i \quad\mathrm{and}\quad
E^a_i = p\, V_o^{-2/3}\,\sqrt{{}^o\!q}\,\, {}^o\!e^a_i,
\end{align} where
$({}^o\!\omega_a^i, {}^o\!e^a_i)$ are a set of orthonormal co-triads
and triads compatible with ${{}^o\!q}_{ab}$ and adapted to the edges
of the elementary cell ${\cal V}$. In terms of $p$, the physical
triad and cotriad are given by
\begin{align}
\label{tri-cotri} e^a_i&=\mathrm{sgn(p)}\,|p|^{-1/2}\,V_o^{1/3}\,
{}^o\!e^a_i,\nonumber\\
e^i_a&=\mathrm{sgn(p)}\,|p|^{1/2}\,V_o^{-1/3}\, {}^o\!\omega^i_a.
\end{align}
The basic (nonvanishing) Poisson bracket is given by
\begin{align}
\label{bracket} \{c,\, p\} = \frac{\kappa\gamma}{3},
\end{align}
where $\kappa=8\pi G$ ($G$ is the Newton's constant) and $\gamma$ is
the Barbero-Immirzi parameter.

To pass to the quantum theory, one constructs a kinematical Hilbert
space ${\cal
H}^{\mathrm{grav}}_{\mathrm{kin}}=L^2(\mathbb{R}_{\mathrm{Bohr}},{\mathrm{d}}\mu_{\mathrm{Bohr}})$,
where $\mathbb{R}_{\mathrm{Bohr}}$ is the Bohr compactification of
the real line and ${\mathrm{d}}\mu_{\mathrm{Bohr}}$ is the Haar
measure on it \cite{math}. The abstract $*$-algebra represented on
the Hilbert space is based on the holonomies  of the connection
$A^i_a$. In the Hamiltonian constraint of LQG, the gravitational
connection $A^i_a$ appears through its curvature $F^i_{ab}$. Since
there exists no operator corresponding to $c$, only holonomy
operators are well defined. Hence one is led to express the
curvature in terms of holonomies. Similarly, in the improved
dynamics setting of LQC \cite{aps}, to express the curvature one
employed the holonomies
\begin{align}
\label{hol} h_i^{(\bar{\mu})}:=\cos\frac{\bar{\mu}
c}{2}\,\mathbb{I}+2\sin\frac{\bar{\mu} c}{2}\,\tau_i
\end{align}
along an edge parallel to the triad ${}^o\!e^a_i$ of length
$\bar{\mu}\sqrt{|p|}\equiv D$, where $D$ is a constant, with respect
to the physical metric $q_{ab}$, where $\mathbb{I}$ is the identity
$2\times2$ matrix and $\tau_i=-i\sigma_i/2$ ($\sigma_i$ are the
Pauli matrices). Thus, the elementary variables could be taken as
the functions $\exp(i\bar{\mu} c/2)$ and the physical volume
$V=|p|^{3/2}$ of the cell, which have unambiguous operator analogs.
It is convenient to work with the $v$-representation. In this
representation, states $|v\rangle$, constituting an orthonormal
basis in ${\cal H}^{\mathrm{grav}}_{\mathrm{kin}}$, is more directly
adapted to the volume operator $\hat{V}$ as
\begin{align}
\hat{V}|v\rangle=\left(\frac{8\pi\gamma\ell_p^2}{6}\right)^{3/2}\frac{|v|}{L'}|v\rangle,
\end{align}
where $\ell_p^2=G\hbar$ and
\begin{align}
L'=\frac{4}{3}\sqrt{\frac{\pi\gamma\ell_p^2}{3D}}.
\end{align}
The action of $\widehat{\exp(i\bar{\mu}c/2)}$ is given by
\begin{align}
\widehat{\exp(i\bar{\mu}c/2)}|v\rangle=|v+1\rangle.
\end{align}

Now let us consider the gravitational field coupled with a massless
scalar field $\phi$. The Hamiltonian of the matter field is given by
$H_{\phi} = |p|^{-3/2}\, p_\phi^2/2$, where $p_\phi$ denotes the
momentum of $\phi$. Hence the total Hamiltonian constraint is given
by $H=H_{\mathrm{grav}}+H_\phi$. The basic Poisson bracket for the
matter field is given by
\begin{align}
\{\phi,p_\phi\}=1,
\end{align}
The Hamiltonian evolution equations for the matter field read
\begin{align}
\dot{p}_\phi&=\{p_\phi,H_{\phi}\}=0\Longrightarrow p_\phi=\mathrm{constant},\nonumber\\
\dot{\phi}&=\{\phi,H_{\phi}\}=\frac{p_\phi}{|p|^{3/2}},
\end{align}
which show that $\phi$ is monotonic function of the time parameter.
So the scalar field can be regarded as internal time. To quantize
the matter field, we can choose the standard Schr\"{o}dinger
representation for scalar field. The kinematical Hilbert space, ${\cal
H}_{\phi}=L^2(\mathbb{R},\mathrm{d}\phi)$, is
the space of the square integrable functions on $\mathbb{R}$ endowed
with the Lebesgue measure. Hence the kinematical
Hilbert space of the gravitational field coupled with the scalar
field is ${\cal H}_{\mathrm{kin}}={\cal
H}^{\mathrm{grav}}_{\mathrm{kin}}\otimes{\cal H}_{\phi}$. The
elementary operators of the scalar field are defined by:
\begin{align}
(\hat{\phi}\Psi)(v,\phi)&:=\phi\Psi(v,\phi),\nonumber\\
(\hat{p}_\phi\Psi)(v,\phi)&:=-i\hbar\frac{d}{d\phi}\Psi(v,\phi)\quad\quad\forall
\Psi(v,\phi)\in{\cal H}_{\mathrm{kin}}.
\end{align}
In the following sections, we will construct two different
Hamiltonian operators including the Lorentz term in the above
quantum kinematic framework. Their classical limits are confirmed by
calculating the expectation values of these new Hamiltonian
operators with respect to suitable semiclassical states. By this
approach we also obtain the effective descriptions of quantum
dynamics in both cases. In the end we will summarize the results and
discuss some of their ramifications.

\section{Alternative regularized Hamiltonian constraints}
Because of homogeneity, we can assume that the lapse
$N$ is constant and from now onwards set it to be one. The
gravitational Hamiltonian constraint is given by
\begin{align}
\label{HEL} H_{\mathrm{grav}}&=\int_{\cal
V}d^3x\frac{E^{aj}E^{bk}}{2\kappa\sqrt{\det(q)}}\big[\epsilon_{ijk}F_{ab}^i
-2(1+\gamma^2)K^j_{[a}K^k_{b]}\big]\nonumber\\
 &\equiv H^E(1)-2(1+\gamma^2)\,T(1).
\end{align}
The symmetry-reduced expressions read
\begin{align}
\label{redH} &H^E(1)=\int_{\cal
   V}d^3x\frac{E^{aj}E^{bk}}{2\kappa\sqrt{\det(q)}}\epsilon_{ijk}F_{ab}^i=\frac{3}{\kappa}\;c^2\sqrt{|p|},\nonumber\\
&T(1)=\int_{\cal
       V}d^3x\frac{E^{aj}E^{bk}}{2\kappa\sqrt{\det(q)}}K^j_{[a}K^k_{b]}=
       \frac{3}{2\kappa\gamma^2}\,c^2\,\sqrt{|p|},\nonumber\\
&H_{\mathrm{grav}}=H^E(1)-2(1+\gamma^2)\,T(1)=-\frac{3}{\kappa\gamma^2}c^2\sqrt{|p|}.
\end{align}
For the spatially flat isotropic model, one has classically the
identity
\begin{align}
 K^i_a=\frac{1}{\gamma}A^i_a\label{A}.
\end{align}
Hence one obtains the classically equivalent expression of Eq.
(\ref{HEL}) as
\begin{align}
\label{dmy2} H^S_{\mathrm{grav}}=H^E(1)-2(1+\gamma^2)T_S(1),
\end{align}
where
\begin{align}
T_S(1)=\frac{1}{\gamma^2}\int_{\cal
       V}d^3x\frac{E^{aj}E^{bk}}{2\kappa\sqrt{\det(q)}}A^j_{[a}A^k_{b]}\label{TS}.
\end{align}
Both Eqs. (\ref{HEL}) and (\ref{dmy2}) are alternative expressions
of the Hamiltonian constraint classically equivalent to the one
currently employed in the literatures (see e.g. \cite{aps}). Since
the expressions (\ref{HEL}) and (\ref{dmy2}) inherit more features
of the Hamiltonian constraint in the full theory, we will take them
separately as the starting-points of our quantization. To this aim,
the first step is to give their regularized expressions which would
be suitable for quantization. Note that the improved quantum
operator representing the Euclidean Hamiltonian constraint $H^E(1)$
was first introduced in \cite{aps}, and its regularized formulation
reads
\begin{align}
\label{HeR}
H^{E,\bar{\mu}}(1)=&\frac{2\,\mathrm{sgn}(p)}{\kappa^2\gamma\bar{\mu}^3}\;
 \epsilon^{ijk}\mathrm{Tr}\Big(h_i^{(\bar{\mu})}h_j^{(\bar{\mu})}{h_i^{(\bar{\mu})}}^{-1}
 {h_j^{(\bar{\mu})}}^{-1}\nonumber\\
 &\quad\quad\quad\times
 h_k^{(\bar{\mu})}\{{h_i^{(\bar{\mu})}}^{-1},V\}\Big).
\end{align}
Now our task is to give the regularized formulations of the Lorentz
terms $T_S(1)$ and $T(1)$ in Eqs. (\ref{dmy2}) and (\ref{HEL})
respectively. Let us first deal with the symmetry-reduced Lorentzian
term $T_S(1)$. By Thiemann's trick \cite{thiem,ThiemQ},
\begin{align}
\label{idEE}
\frac{\epsilon_{ijk}E^{aj}E^{bk}}{\sqrt{\det(q)}}=\frac{2}{\kappa\gamma}\;\tilde{\epsilon}^{abc}\{A^i_c,V\},
\end{align}
where $\tilde{\epsilon}^{abc}$ is the Levi-Civita density, Eq.
(\ref{TS}) can be written as
\begin{align}
\label{Ts1} T_S(1)&=-\frac{2}{\kappa^2\gamma^3}\int_{\cal
V}d^3x\tilde{\epsilon}^{abc}\mathrm{Tr}(A_aA_b\{A_c,V\})\nonumber\\
&=-\frac{2\mathrm{sgn}(p)}{\kappa^2\gamma^3}\epsilon^{ijk}\mathrm{Tr}\left(c\tau_ic\tau_j\{c\tau_k,V\}\right).
\end{align}
Moreover it is easy to show that
\begin{align}
&c\tau_i=\lim_{\bar{\mu}\rightarrow0}\frac{1}{2\bar{\mu}}\left[h^{(\bar{\mu})}_i
-{h^{(\bar{\mu})}_i}^{-1}\right]\label{Ah},\\
\label{Pid}
&\{c\tau_k,V\}=-\frac{1}{\bar{\mu}}\;h_k^{(\bar{\mu})}\{{h_k^{(\bar{\mu})}}^{-1},V\}.
\end{align}
Hence Eq. (\ref{Ts1}) can be written as
\begin{align}
\label{Ts} T_S(1)&=\lim_{\bar{\mu}\rightarrow0}T^{\bar{\mu}}_S(1),
\end{align}
where
\begin{align}
\label{Tsr}
T^{\bar{\mu}}_S(1)&=\frac{\mathrm{sgn}(p)}{2\kappa^2\gamma^3\bar{\mu}^3}\epsilon^{ijk}\mathrm{Tr}\Bigg(
\left[h^{(\bar{\mu})}_i-{h^{(\bar{\mu})}_i}^{-1}\right]\nonumber\\
&\quad\quad\left[h^{(\bar{\mu})}_j-{h^{(\bar{\mu})}_j}^{-1}\right]h_k^{(\bar{\mu})}\{{h_k^{(\bar{\mu})}}^{-1},V\}\Bigg)
\end{align}
Putting Eqs. (\ref{HeR}) and (\ref{Tsr}) together, we obtain the
regularized Hamiltonian constraint corresponding to (\ref{dmy2}) as
\begin{align}
\label{Hs}
H^{S,\bar{\mu}}_{\mathrm{grav}}=H^{E,\bar{\mu}}(1)-2(1+\gamma^2)T^{\bar{\mu}}_S(1).
\end{align}

Let us now turn to the original Lorentz term in Eq. (\ref{HEL}):
\begin{align}
T(1)=\int_{\cal
       V}d^3x\frac{E^{aj}E^{bk}}{2\kappa\sqrt{\det(q)}}K^j_{[a}K^k_{b]}\label{T1}.
\end{align}
Though this term was considered in some early literature \cite{Boj},
here we will treat it in the new improved quantization framework
\cite{aps}. As in the full theory \cite{thiem,ThiemQ}, the extrinsic
curvature can be written as
\begin{align}
\label{id22}
K^i_a=\frac{1}{\kappa\gamma}\{A_a^i,K\}=\frac{1}{\kappa\gamma^3}\{A_a^i,\{H^E(1),V\}\},
\end{align}
where
\begin{align}
\label{KrEx} K=\int_{\cal V} d^3xK^i_aE^a_i=\frac{3}{\gamma}\,cp
\end{align}
is the integrated trace of
$K^i_a$. Hence Eq. (\ref{T1}) can be reexpressed as
\begin{align}
\label{Tf1} T(1)&=-\frac{2}{\kappa^4\gamma^3}\int_{\cal
V}d^3x\;\tilde{\epsilon}^{abc}\mathrm{Tr}\left(\{A_a,K\}\{A_b,K\}\{A_c,V\}\right)\nonumber\\
&=-\frac{2\,\mathrm{sgn}(p)}{\kappa^4\gamma^3}\,\epsilon^{ijk}\mathrm{Tr}\left(\{c\,\tau_i,K\}\{c\,\tau_j,K\}
\{c\,\tau_k,V\}\right).
\end{align}
Moreover, we have the following identity
\begin{align}
\label{ident1}
\{c\tau_i,K\}&=-\frac{2}{3\bar{\mu}}h_i^{(\bar{\mu})}\{{h_i^{(\bar{\mu})}}^{-1},K\}.
 \end{align}
Using the identities ({\ref{Pid}}) and (\ref{ident1}), Eq.
(\ref{Tf1}) can be written as
\begin{align}
\label{Tf} T(1)&=\lim_{\bar{\mu}\rightarrow0}T^{\bar{\mu}}_F(1),
\end{align}
where
\begin{align}
\label{TR}
T^{\bar{\mu}}_F(1)&=\frac{8\,\mathrm{sgn}(p)}{9\kappa^4\gamma^7\bar{\mu}^3}\epsilon^{ijk}\mathrm{Tr}
\Big(h_i^{(\bar{\mu})}\left\{{h_i^{(\bar{\mu})}}^{-1},\{H^{E,\bar{\mu}}(1),V\}\right\}\nonumber\\
 &\quad\times h_j^{(\bar{\mu})}\left\{{h_j^{(\bar{\mu})}}^{-1},\{H^{E,\bar{\mu}}(1),V\}\right\}\nonumber\\
 &\quad\times
 h_k^{(\bar{\mu})}\{{h_k^{(\bar{\mu})}}^{-1},V\}\Big).
\end{align}
Putting Eqs. (\ref{HeR}) and (\ref{TR}) together, we obtain the
regularized Hamiltonian constraint corresponding to (\ref{HEL}) as
\begin{align}
\label{Hf}
H^{F,\bar{\mu}}_{\mathrm{grav}}=H^{E,\bar{\mu}}(1)-2(1+\gamma^2)T^{\bar{\mu}}_F(1).
\end{align}

\section{Alternative Hamiltonian constraint operators}
Since both regularized Hamiltonian constraints (\ref{Hs}) and
(\ref{Hf}) are now expressed in terms of elementary variables and
their poisson brackets, which have unambiguous quantum analogs, it
is straightforward to write down the quantum operators
$\hat{H}^{S,\bar{\mu}}_{\mathrm{grav}}$ and
$\hat{H}^{F,\bar{\mu}}_{\mathrm{grav}}$. However, the limit
$\bar{\mu}\rightarrow0$ of these operators do not exist, not only
for the Euclidean term $\hat{H}^{E,\bar{\mu}}(1)$, but also for the
Lorentzian term $\hat{T}^{\bar{\mu}}_S(1)$ or
$\hat{T}^{\bar{\mu}}_F(1)$. In fact, even in the full theory, there
are no local operators representing connections and curvatures. To
get unambiguous operators, one should have recourse to the area gap
as in the improved scheme \cite{aps}, where $\bar{\mu}$ is given by
\begin{align}
\bar{\mu}^2|p|=\Delta,
\end{align}
here $\Delta=4\sqrt{3}\pi\gamma\ell^2_p$ is a minimum nonzero
eigenvalue of the area operator \cite{ash-r}. The Euclidean
Hamiltonian constraint operator $\hat{H}^E(1)$ corresponding to
(\ref{HeR}) is given in the improved scheme by \cite{aps}
\begin{align}
\hat{H}^E(1)&=-\frac{\gamma^2}{2\kappa}\sin(\bar{\mu}c)\Big[\frac{24\,i\,\mathrm{sgn}(v)}{\kappa\hbar\gamma^3
\bar{\mu}^3}\Big(\sin\left(\frac{\bar{\mu}c}{2}\right)\hat{V}
 \cos\left(\frac{\bar{\mu}c}{2}\right)\nonumber\\
 &\quad-\cos\left(\frac{\bar{\mu}c}{2}\right)\hat{V}\sin\left(\frac{\bar{\mu}c}{2}\right)\Big)\Big]\sin(\bar{\mu}c).
 \label{Heucl}
\end{align}
where, for clarity, we have suppressed hats over the operators
$\sin(\bar{\mu} c/2)$, $\cos(\bar{\mu} c/2)$ and
$\mathrm{sgn}(v)/\bar{\mu}^3$. In the $v$-representation where
\begin{equation}
v:=\frac{\mathrm{sgn}(p)|p|^{3/2}}{2\pi\gamma\ell_p^2\sqrt{\Delta}},
\end{equation}
$\hat{H}^E(1)$ acts on the basis $|v\rangle$ of ${\cal
H}^{\mathrm{grav}}_{\mathrm{kin}}$ as
\begin{align}
\hat{H}^E(1) \, |v\rangle &=-\frac{\gamma^2}{2\kappa}\bigg[f_+(v)
\,|v + 4\rangle +
f_o(v) \, |v\rangle \nonumber\\
&\quad\quad\quad\quad + f_-(v) \, |v - 4\rangle\bigg],
\end{align}
where
\begin{align}
f_+(v)&=\frac{27}{16}\sqrt{\frac{8\pi}{6}}\frac{L\ell_p}{\gamma^{3/2}}\big(|v+3|-|v+1|\big)(v+2),\nonumber\\
f_-(v)&=f_+(v-4),\quad f_o(v)=-f_+(v)-f_-(v),
\end{align}
here
\begin{align}
L=\frac{4}{3}\sqrt{\frac{\pi\gamma\ell_p^2}{3\Delta}}.
\end{align}

Now we turn to the Lorentz part. The regularized expression
(\ref{Tsr}) enables us to define a self-adjoint operator on
$\mathcal{H}^{\mathrm{grav}}_{\mathrm{kin}}$ in the improved scheme
corresponding to (\ref{TS}) as
\begin{align}
\label{Tsop} \hat{T}_S(1)&=-\sin\frac{\bar{\mu}
    c}{2}\Big[\frac{24i\,\mathrm{sgn}(v)}{\kappa^2\hbar\gamma^3
    \bar{\mu}^3}\Big(\sin\frac{\bar{\mu}
    c}{2}\hat{V}\cos\frac{\bar{\mu}
    c}{2}\nonumber\\
    &\quad\quad\quad\quad-\cos\frac{\bar{\mu}
    c}{2}\hat{V}\sin\frac{\bar{\mu}
    c}{2}\Big)\Big]\sin\frac{\bar{\mu}
    c}{2}.
\end{align}
Its action on the basis $|v\rangle$ reads
\begin{align}
\label{Tsac}
\hat{T}_S(1)|v\rangle=S_+(v)|v+2\rangle+S_o(v)|v\rangle+S_-(v)|v-2\rangle,
\end{align}
where
\begin{align}
S_+(v)&=-\frac{27}{16}\sqrt{\frac{8\pi}{6}}\frac{L\ell_p}{\kappa\gamma^{3/2}}(v+1)\big(|v+2|-|v|\big),\nonumber\\
S_-(v)&=S_+(v-2),\quad S_o(v)=-S_+(v)-S_-(v).
\end{align}
Hence we obtain the action of the Hamiltonian constraint operator
corresponding to (\ref{dmy2}) on $|v\rangle$ as
\begin{align}
\label{HSact}
\hat{H}^S_{\mathrm{grav}}|v\rangle=&\hat{H}^E_{\mathrm{grav}}(1)|v\rangle-2(1+\gamma^2)\hat{T}_S(1)|v\rangle\nonumber\\
 =&f'_+(v)|v+4\rangle+S'_+(v)|v+2\rangle\nonumber\\
  &+\left[f'_o(v)+S'_o(v)\right]|v\rangle\nonumber\\
  &+S'_-(v)|v-2\rangle+f'_-(v)|v-4\rangle,
\end{align}
where $f'_*(v)=-\frac{\gamma^2}{2\kappa}f_*(v)$,
$S'_*(v):=-2(1+\gamma^2)S_*(v)$, here $*=+, -, o$.

On the other hand, the regularized expression (\ref{TR}) enables us
to define the other operator on
$\mathcal{H}^{\mathrm{grav}}_{\mathrm{kin}}$ corresponding to
(\ref{T1}) as
\begin{align}
\label{Tfop}
\hat{T}_F(1)&=-\frac{96i}{9\kappa^4\gamma^7\hbar^5}\left(\sin\frac{\bar{\mu}
       c}{2}\hat{B}\cos\frac{\bar{\mu}
       c}{2}-\cos\frac{\bar{\mu}
       c}{2}\hat{B}\sin\frac{\bar{\mu}
       c}{2}\right)\nonumber\\
       &\;\;\;\;\;\;\times\,\left[\frac{\mathrm{sgn}(v)}{\bar{\mu}^3}\,\left(\sin\frac{\bar{\mu}
       c}{2}\hat{V}\cos\frac{\bar{\mu}
       c}{2}-\cos\frac{\bar{\mu}
       c}{2}\hat{V}\sin\frac{\bar{\mu}
       c}{2}\right)\right]\nonumber\\
       &\quad\times\left(\sin\frac{\bar{\mu}
       c}{2}\hat{B}\cos\frac{\bar{\mu}
       c}{2}-\cos\frac{\bar{\mu}
       c}{2}\hat{B}\sin\frac{\bar{\mu}
       c}{2}\right),
\end{align}
where
\begin{align}
\hat{B}&\equiv[\hat{H}^E(1),\hat{V}].
\end{align}
It is easy to see from Eq. (\ref{Tfop}) that $\hat{T}_F(1)$ is a
symmetric operator. Its action on $|v\rangle$ reads
\begin{align}
\label{Tfac} \hat{T}_F(1)|v\rangle
 &=\frac{\sqrt{6}}{2^{8}\times 3^3}\,\frac{\gamma^{3/2}}{\kappa^{3/2}\hbar^{1/2}}\,\frac{1}{L}\Big(F_+(v)|v
 +8\rangle+F_o(v)|v\rangle\nonumber\\
 &\quad+F_-(v)|v-8\rangle\Big),
\end{align}
where
\begin{align}
F_+(v)&=\Big[M_v(1,5)f_+(v+1)-
     M_v(-1,3)f_+(v-1)\Big]\nonumber\\
       &\quad\times(v+4)M_v(3,5)\nonumber\\
       &\quad\times\Big[M_v(5,9)f_+(v+5)-M_v(3,7)f_+(v+3)\Big],\nonumber\\
F_-(v)&=\Big[M_v(1,-3)f_-(v+1)-M_v(-1,-5)f_-(v-1)\Big]\nonumber\\
       &\quad\times(v-4)M_v(-5,-3)\nonumber\\
       &\quad\times\big[M_v(-3,-7)f_-(v-3)-M_v(-5,-9)f_-(v-5)\big],\nonumber\\
F_o(v)&=\Big[M_v(1,5)f_+(v+1)-M_v(-1,3)f_+(v-1)\Big]\nonumber\\
       &\quad\times(v+4)M_v(3,5)\nonumber\\
       &\quad\times\Big[M_v(5,1)f_-(v+5)-M_v(3,-1)f_-(v+3)\Big]\nonumber\\
       &+\Big[M_v(1,-3)f_-(v+1)-M_v(-1,-5)f_-(v-1)\Big]\nonumber\\
       &\quad\times(v-4)M_v(-5,-3)\nonumber\\
       &\quad\times\Big[M_v(-3,1)f_+(v-3)-M_v(-5,-1)f_+(v-5)\big],
\end{align}
here
\begin{align}
M_v(a,b):=|v+a|-|v+b|.
\end{align}
Hence the action of the Hamiltonian constraint operator
corresponding to (\ref{HEL}) on $|v\rangle$ is given by
\begin{align}
\label{HFact}
\hat{H}^F_{\mathrm{grav}}|v\rangle=&\hat{H}^E_{\mathrm{grav}}(1)|v\rangle-2(1+\gamma^2)\hat{T}_F(1)|v\rangle\nonumber\\
 =&F'_+(v)|v+8\rangle+f'_+(v)|v+4\rangle\nonumber\\
  &+\left[F'_o(v)+f'_o(v)\right]|v\rangle\nonumber\\
  &+f'_-(v)|v-4\rangle+F'_-(v)|v-8\rangle,
\end{align}
where $F'_*(v):=-2(1+\gamma^2)\frac{\sqrt{6}}{2^{8}\times
3^3}\,\frac{\gamma^{3/2}}{\kappa^{3/2}\hbar^{1/2}}\,\frac{1}{L}F_*(v)$,
here $*=+, -, o$.

Thus, both of the new proposed Hamiltonian constraint operators in
Eqs. (\ref{HSact}) and (\ref{HFact}) are also difference operators
with constant steps in eigenvalues of the volume operator $\hat{V}$.
But they contain more terms with steps of different size comparing
to the operator (\ref{Heucl}).

The Hamiltonian constraint of the scalar field has been quantized in
the literatures as \cite{aps,math}
\begin{equation}
 \hat{H}_{\phi}=
\frac{1}{2}\,\widehat{|p|^{-3/2}}\, \widehat{p_\phi^2},
\end{equation}
where the action of $\widehat{|p|^{-3/2}}$ on $|v\rangle$ reads
\begin{align}
\widehat{|p|^{-3/2}}|v\rangle=&\left(\frac{3}{2}\right)^3\left(\frac{6}{\kappa\hbar\gamma}\right)^{3/2}L|v|\nonumber\\
&\times\big||v+1|^{1/3}-|v-1|^{1/3}|^3\big|v\rangle.
\end{align}
Thus we obtain alternative total Hamiltonian operators respectively
as
\begin{align}
 &\hat{H}_S=\hat{H}^S_{\mathrm{grav}}+\hat{H}_\phi\label{HSo},\\
 &\hat{H}_F=\hat{H}^F_{\mathrm{grav}}+\hat{H}_\phi\label{HFo}.
\end{align}
Note that in both quantum dynamics the scalar field $\phi$ can be
used as emergent time. But the different expressions of
gravitational Hamiltonian operators may lead to different physics,
which may be examined in some aspects. We will consider their
classical limit and effective dynamics in next section.

\section{Classical limit and effective dynamics}
It has been shown in \cite{ta,dmy1} that the improved Hamiltonian
constraint operator constructed in \cite{aps} has the correct
classical limit. In this section, we will show that the two
Hamiltonian constraint operators constructed in this paper also have
the correct classical limit. Moreover, the effective Hamiltonian
incorporating higher order quantum corrections can also be obtained.
In order to do the semiclassical analysis, it is convenient to
introduce a new variable:
\begin{align}
b:=\frac{\sqrt{\Delta}}{2}\frac{c}{\sqrt{|p|}}
\end{align}
conjugate to $v$ with the Poisson bracket $\{b,v\}=1/\hbar$. Then
the classical Hamiltonian constraint in FRW model can be written as
\begin{align}
\label{Hbv} H&=H_{\rm grav}+H_\phi\nonumber\\
&=-\frac{3^2\sqrt{6}}{2}\frac{\hbar^{1/2}}{\gamma^{3/2}\kappa^{1/2}}L\,|v|\,b^2+
\left(\frac{\kappa\gamma\hbar}{6}\right)^{3/2}\frac{|v|}{L}\;\rho,
\end{align}
where
\begin{align}
\rho=\f{1}{2}\left(\f{6}{\kappa\gamma
\hbar}\right)^{3}\left(\f{L}{|v|}\right)^2p_{\phi}^2
\end{align}
is the energy density of the scalar field. Let us first consider the
gravitational part. Since there are uncountable basis vectors, the
natural Gaussian semiclassical states live in the algebraic dual
space of some dense set in ${\cal H}^{\rm{grav}}_{\rm{kin}}$. A
semiclassical state $(\Psi_{(b_o,v_o)}|$ peaked at a point
$(b_o,v_o)$ of the gravitational classical phase space reads:
\begin{align}
(\Psi_{(b_o,\, v_o)}|=\sum_{v\in
\mathbb{R}}e^{-[(v-v_o)^2/2d^2]}e^{i\,b_o(v-v_o)}(v|,\label{coh}
\end{align}
where $d$ is the characteristic ``width'' of the coherent state. For
practical calculations, we use the shadow of the semiclassical state
$(\Psi_{(b_o,v_o)}|$ on the regular lattice with spacing 1
\cite{shad}, which is given by
\begin{align}
|\Psi\rangle \,=\, \sum_{n\in\Z}\,\left[e^{-(\epsilon^2/2)(n-N)^2}\,
\,e^{-i\, (n-N)b_o}\right] |n\rangle ,\label{shad}
\end{align}
where $\epsilon = 1/d$ and we choose $v_o=N\in\mathbb{Z}$. Since we
consider large volumes and late times, the relative quantum
fluctuations in the volume of the universe must be very small.
Therefore we have the restrictions: $ {1}/{N}\ll\epsilon\ll1$ and $
b_o\ll 1 $. One can check that the state (\ref{coh}) is sharply
peaked at $(b_o,v_o)$ and the fluctuations are within specified
tolerance \cite{dmy1,ta}. The semiclassical state of matter part is
given by the standard coherent state
\begin{align}
(\Psi_{(\phi_o,p_{\phi})}|=\int
{\mathrm{d}}\phi\,e^{-[(\phi-\phi_o)^2/2\sigma^2]}\,e^{{i
p_\phi}(\phi-\phi_o)/\hbar}(\phi|\label{matterstate},
\end{align}
where $\sigma$ is the width of the Gaussian. Thus the whole
semiclassical state reads $(\Psi_{(b_o,\,
v_o)}|\bigotimes(\Psi_{(\phi_o,p_{\phi})}|$.

\begin{figure}[!htb]
    \includegraphics[width=0.45\textwidth,angle=0]{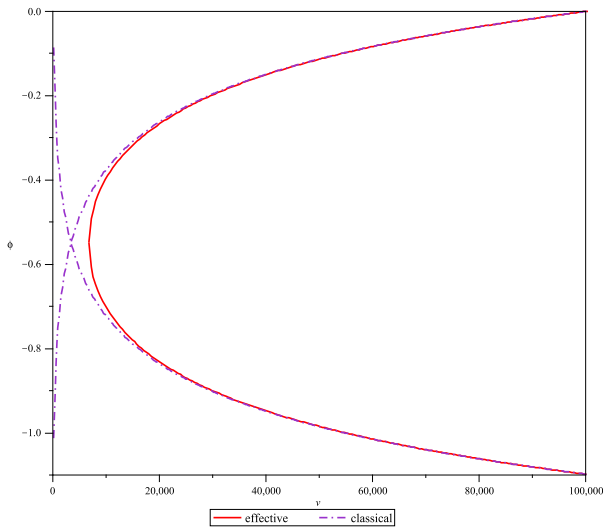}
    \caption{The effective dynamics represented by the observable
    $v|_{\phi}$ are compared to classical trajectories. In this
    simulation, the parameters were:
    $G=\hbar=1\,$, $p_{\phi}=10\,000,$
    $\epsilon=0.001\,,\sigma=0.01$ with initial data
    $v_o=100\,000\,$.}
    \label{full:bounce}
\end{figure}

The task is to use this semiclassical state to calculate the
expectation values of the two Hamiltonian operators in Eqs.
(\ref{HSo}) and (\ref{HFo}) to a certain accuracy. In the
calculation, one gets the expression with the absolute values, which
is not analytical. To overcome the difficulty we separate the
expression into a sum of two terms: one is analytical and hence can
be calculated straightforwardly, while the other becomes
exponentially decayed out. We first see the expectation value of the
Hamiltonian operator $\hat{H}_F$, which inherits more features of
the full theory. The expectation values of the two terms of
$\hat{H}^F_{\mathrm{grav}}$ in Eq. (\ref{HFact}) are respectively
calculated as
\begin{align}
\langle\hat{H}^E(1)\rangle
 &=\frac{3^2\sqrt{6}}{2^3}\,\frac{\gamma^{1/2}\,\hbar^{1/2}}{\kappa^{1/2}}L\,|v_o|\,\Big[e^{-4\epsilon^2}
 \sin^2(2b_o)\nonumber\\
 &\quad\quad+\frac{1}{2}(1-e^{-4\epsilon^2})\Big]+O(e^{-N^2\epsilon^2}),\nonumber\\
\langle\hat{T}_F(1)\rangle=&\frac{3^2\sqrt{6}}{2^6}\,\frac{\hbar^{1/2}}{\gamma^{3/2}\,\kappa^{1/2}}L\,|v_o|\,
\Big[e^{-16\epsilon^2}\sin^2(4b_o)\nonumber\\
&+\frac{1}{2}(1-e^{-16\epsilon^2})\Big]+O(e^{-N^2\epsilon^2}).
\end{align}
In the calculation of $\langle\hat{H}_{\phi}\rangle$, one has to
calculate the expectation value of the operator
$\widehat{|p|^{-3/2}}$, which is given by \cite{dmy1}:
\begin{align}
\langle\widehat{|p|^{-3/2}}\rangle=&\left(\frac{6}{8 \pi \gamma
\ell_p^2}\right)^{3/2}\,\frac{L}{N}\Big[1+\frac{1}{2N^2\epsilon^2}+\frac{5}{9N^2}\nonumber\\
&+O(1/N^4\epsilon^4)\Big]+O\big(e^{-N^2\epsilon^2}\big)
+O\big(e^{-\pi^2/\epsilon^2}\big).\nonumber
\end{align}
For clarity, we will suppress the label $o$ in the following.
Collecting these results we can obtain an effective Hamiltonian with
the relevant quantum corrections of order
$\epsilon^2,1/v^2\epsilon^2,\hbar^2/\sigma^2p^2_\phi$ as:
\begin{align}
\label{effHf} {\cal
H}^F_{\mathrm{eff}}&=-\frac{3^2\sqrt{6}}{2^3}\,\frac{\hbar^{1/2}}{\gamma^{3/2}\,\kappa^{1/2}}L\,|v|\nonumber\\
&\quad\times\Big[\sin^2(2b)\big[1-(1+\gamma^2)\sin^2(2b)\big]+2\epsilon^2\Big]
\nonumber\\
&\quad+\left(\frac{\kappa\gamma\hbar}{6}\right)^{3/2}\frac{|v|}{L}\rho\left(1+\frac{1}{2|v|^2\epsilon^2}+
\frac{\hbar^2}{2\sigma^2p_\phi^2}\right).
\end{align}
Hence the classical constraint (\ref{Hbv}) is reproduced up to small
quantum corrections and therefore the Hamiltonian operator
$\hat{H}_F$ has correct classical limit. We can further obtain the
Hamiltonian evolution equation of $v$ by taking its Poisson bracket
with ${\cal H}^F_{\mathrm{eff}}$ as
\begin{align}
\label{dotv}
\dot{v}_F=&3|v|\sqrt{\frac{\kappa}{3}\rho_c}\,\sin(2b)\cos(2b)\Big[1-2(1+\gamma^2)\sin^2(2b)\Big],
\end{align}
where $\rho_c=3/(\kappa\gamma^2\Delta)$. The vanishing of the
effective Hamiltonian constraint (\ref{effHf}) gives
\begin{align}
\sin^2(2b)&\big[1-(1+\gamma^2)\sin^2(2b)\big]\nonumber\\
&=\frac{\rho}{\rho_c}\left(1+\frac{1}{2|v|^2\epsilon^2}+\frac{\hbar^2}{2\sigma^2p_\phi^2}\right)-2\epsilon^2\label{sin2}.
\end{align}
For the classical region, $b\ll1$ and $\rho\ll\rho_c$, we have from
Eq. (\ref{sin2})
\begin{align}
\sin^2(2b)&=\frac{1-\sqrt{1-\chi_F}}{2(1+\gamma^2)},\label{sin22}
\end{align}
where
\begin{align}
\chi_F&=4(1+\gamma^2)\left[\frac{\rho}{\rho_c}\left(1+\frac{1}{2|v|^2\epsilon^2}+
\frac{\hbar^2}{2\sigma^2p_\phi^2}\right)-2\epsilon^2\right].\label{chif}
\end{align}
The modified Friedmann equation can then be derived as
\begin{align}
\label{HubF}
H_F^2&=\left(\frac{\dot{v}_F}{3v}\right)^2\nonumber\\
&=\frac{\kappa}{3}\frac{\rho_c}{4(1+\gamma^2)^2}\left(1-\sqrt{1-\chi_F}\,\right)\left(1+2\gamma^2
+\sqrt{1-\chi_F}\,\right)\nonumber\\
&\quad\times\left(1-\chi_F\right).
\end{align}
It is easy to see that if one neglects the small quantum corrections
in the classical region, $\chi_F\ll1$ for $\rho\ll\rho_c$, one gets
\begin{align}
H_F^2
\approx\frac{\kappa}{3}\frac{\rho_c}{4(1+\gamma^2)^2}\frac{1}{2}\chi_F2(1+\gamma^2)\approx\frac{\kappa}{3}\rho,
\end{align}
which reduces to the standard Friedmann equation. However, quantum
geometry effects lead to a modification of the Friedmann equation
especially at the scales when $\rho$ becomes comparable to $\rho_c$.
Remarkable changes to the classical theory happen when the Hubble
parameter in Eq. (\ref{HubF}) vanishes by
\begin{align}
&1-\chi_F=0.
\end{align}
If we consider only the leading order contribution in Eq.
(\ref{chif}), this can happen when
\begin{equation}
\rho=\rho_c/4(1+\gamma^2).
\end{equation}
Thus, when energy density of the scalar field reaches to the leading
order critical energy density $\rho_c^{F}=\rho_c/4(1+\gamma^2)$, the
universe bounces from the contracting branch to the expanding
branch. The quantum bounce implied by (\ref{HubF}) is shown in Fig.
\ref{full:bounce} .

In a similar way, we can calculate the expectation value of the
other Hamiltonian operator (\ref{HSo}) as well. The effective
Hamiltonian corresponding to $\hat{H}_S$ with the relevant quantum
corrections is obtained as
\begin{align}
\label{effHs}
{\cal H}^S_{\mathrm{eff}}&=-\frac{3^2\sqrt{6}}{2}\,\frac{\hbar^{1/2}}{\gamma^{3/2}\,\kappa^{1/2}}L\,|v|\nonumber\\
&\quad\quad\times\left[\sin^2b\big(1+\gamma^2\sin^2b\big)+\frac{1}{2}\epsilon^2\right]\nonumber\\
&\quad+\left(\frac{\kappa\gamma\hbar}{6}\right)^{3/2}\frac{|v|}{L}\rho\left(1+\frac{1}{2|v|^2\epsilon^2}
+\frac{\hbar^2}{2\sigma^2p_\phi^2}\right).
\end{align}
Hence the classical constraint (\ref{Hbv}) is again reproduced up to
small quantum corrections. The corresponding modified Friedmann
equation can then be derived as
\begin{align}
\label{HubS}
H_S^2
 &=\frac{\kappa}{3}\frac{\rho_c}{\gamma^4}(-1+\sqrt{1+\chi_S}\,)(1+2\gamma^2-\sqrt{1+\chi_S})(1+\chi_S),
\end{align}
where
\begin{align}
\chi_S&=\gamma^2\left[\frac{\rho}{\rho_c}\left(1+\frac{1}{2|v|^2\epsilon^2}+\frac{\hbar^2}{2\sigma^2p_\phi^2}\right)
-\frac{1}{2}\epsilon^2\right]\label{chis}.
\end{align}
The Hubble parameter in Eq. (\ref{HubS}) can also vanish when
\begin{align}
&1+2\gamma^2-\sqrt{1+\chi_S}=0.
\end{align}
Thus the quantum dynamics given by $\hat{H}_S$ has qualitatively
similar feature of that given by $\hat{H}_F$. However, there are
quantitative differences between them. For the leading order
effective theory of $\hat{H}_S$, when energy density of the scalar
field reaches to the critical energy density
$\rho_c^S=4(1+\gamma^2)\rho_c$, the universe bounces from the
contracting branch to the expanding branch.

\section{discussion}

We have successfully constructed two versions of Hamiltonian
operator for isotropic LQC in the improved scheme, where the Lorentz
term is quantized in two approaches different from the Euclidean
one. The treatments of the Lorentz term of Hamiltonian in the full
LQG can be properly implemented in one of our constructions, which
inherits more features of the full theory. One of the Hamiltonian
operators is self-adjoint and the other is symmetric. Both of them
are shown to have the correct classical limit by the semiclassical
analysis. Hence the alternative Hamiltonian operators that we
proposed can provide good arenas to test the ideas and constructions
of the quantum dynamics in full LQG. In the spatially flat FRW model
with a massless scalar field, the effective Hamiltonians and
Friedmann equations are derived in both case. Although there are
quantitative differences between the two versions of quantum
dynamics, qualitatively they have the same dynamical features.
Especially, the classical big bang is again replaced by a quantum
bounce in both cases. For instance, in the leading order effective
theory of $\hat{H}_F$, the universe would bounce from the
contracting branch to the expanding branch when the energy density
of scalar field reaches to the critical
$\rho_c^{F}=\rho_c/4(1+\gamma^2)$. Therefore, the key feature of LQC
for the resolution of the big bang singularity is still maintained
for the new quantum dynamics inheriting more features of the full
theory. Recall that the quantum bounce happens at $\rho_c$ for the
quantum dynamics in \cite{aps}. Thus the new quantum dynamics here
lead to some quantitatively different critical energy density for
the bounce.

\begin{figure}[!htb]
    \includegraphics[width=0.45\textwidth,angle=0]{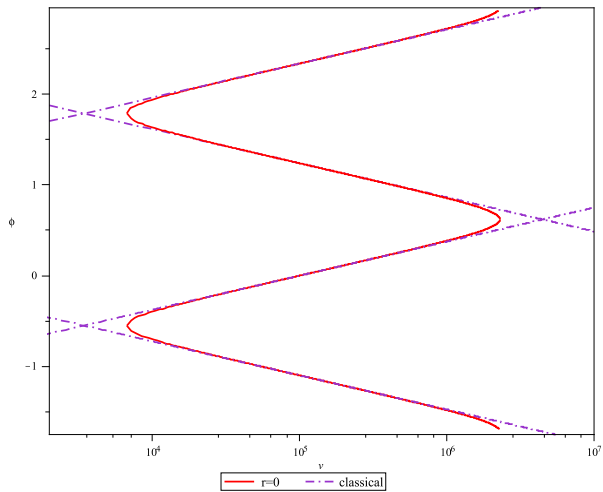}
    \caption{The cyclic model is compared with expanding and contracting classical trajectories.
    In this
    simulation, the parameters were: $G=\hbar=1$\,, $p_{\phi}=10\,000\,,\epsilon=0.001,\sigma=0.01$ with initial data
    $v_o=100\,000$.}
    \label{full:recollapse}
\end{figure}

On the other hand, the discussion in \cite{dmy1} can be carried out
similarly. It is easy to see from Eqs. (\ref{HubF}) and (\ref{HubS})
that the Hubble parameter in both cases may also vanish by the
vanishing of $\chi_F$ and $\chi_S$ respectively, whence the
asymptotic behavior of the quantum geometric fluctuations plays a
key role for the fate of the universe. By the ansatz
$\epsilon=\lambda(r)v^{-r(\phi)}$ with $0\leq r(\phi)\leq1$, Eqs.
(\ref{chif}) and (\ref{chis}) imply that there are great
possibilities for the expanding universe to undergo a recollape in
the future. The recollape can happen provided $0\leq r<1$ in the
large scale limit. Suppose that the semiclassicality of our coherent
state is maintained asymtotically so that the quantum fluctuation
$1/\epsilon$ of $v$ cannot increase as $v$ unboundedly as $v$
approaches infinity. Thus the recollape is in all probability as
viewed from the parameter space of $r(\phi)$. Taking the effective
dynamics of $\hat{H}_F$ as an example, in the scenario when $r=0$
asymtotically, besides the quantum bounce when the matter density
$\rho$ increases to the Planck scale, the universe would also
undergo a recollapse when $\rho$ decreases to
$\rho^{F}_{\mathrm{coll}}\approx8(1+\gamma^2)\epsilon^2\rho^{F}_c$.
Therefore, the quantum fluctuations also lead to a cyclic universe
in this case. The cyclic universe in this effective scenario is
illustrated in Fig. \ref{full:recollapse}.
This amazing possibility that quantum gravity manifests herself in
the large scale cosmology was first revealed in \cite{dmy1}.
Nevertheless, the condition that the semiclassicality is maintained
in the large scale limit has not been confirmed. Hence further
numerical and analytic investigations to the properties of dynamical
semiclassical states in the alternative quantum dynamics are still
desirable. It should be noted that in some simplified completely
solvable models of LQC (see \cite{bojow-r} and \cite{acs}), the
dynamical coherent states could be obtained, where $r(\phi)$
approaches $1$ in the large scale limit. While those treatments lead
to the quantum dynamics different from ours, they raise caveats to
the conjectured recollapse.

To summarize, the quantum dynamics of LQC in the improved scheme is
extended in order to inherit more features from the full LQG. The
key features of LQC in this model, that the big bang singularity is
replaced by a quantum bounce and there are great possibilities for
an expanding universe to recollapse, are robust against the
quantization ambiguities with the extensions. This result further
supports the expectation that the above features of LQC originate
not only in the symmetric model but also from the fundamental LQG.

\section*{ACKNOWLEDGMENTS}

We would like to thank Dah-Wei Chiou for discussions. This work is a
part of project 10675019 supported by NSFC. Y. Ma is grateful to
Thomas Thiemann for the hospitality at AEI and helpful discussions
and acknowledges the financial support from the AEI.

\end{document}